# A novel combination of theoretical analysis and data-driven method for reconstruction of structural defects

Qi Li[1, #], Yihui Da[1, #], Yinghong Zhang[1], Bin Wang[1], Dianzi Liu[2, *], Zhenghua Qian[1, *]

[1]*State Key Laboratory of Mechanics and Control of Mechanical Structures, College of Aerospace Engineering, Nanjing University of Aeronautics and Astronautics, Nanjing, 210016, China*
[2]*School of Engineering, University of East Anglia, UK*

[#] Both authors contributed equally to this manuscript
[*] Corresponding authors: dianzi.liu@uea.ac.uk and qianzh@nuaa.edu.cn

**Abstract:** Ultrasonic guided wave technology has played a significant role in the field of non-destructive testing as it employs acoustic waves that have advantages of high propagation efficiency and low energy consumption during the inspect process. However, theoretical solutions to guided wave scattering problems using assumptions such as Born approximation, have led to the poor quality of the reconstructed results. To address this issue, a novel approach to quantitative reconstruction of defects using the integration of data-driven method with the guided wave scattering analysis has been proposed in this paper. Based on the geometrical information of defects and initial results by the theoretical analysis of defect reconstructions, a deep learning neural network model is built to reveal the physical relationship between defects and the received signals. This data-driven model is then applied to quantitatively assess and characterize defect profiles in structures, reduce the inaccuracy of the theoretical modelling and eliminate the impact of noise pollution in the process of inspection. To demonstrate advantages of the developed approach to reconstructions of defects with complex profiles, numerical examples including basic defect profiles and a defect with the noisy fringe have been examined. Results show that this approach has greater accuracy for reconstruction of defects in structures as compared with the analytical method and provides a valuable insight into the development of artificial intelligence-assisted inspection systems with high accuracy and efficiency in the field of non-destructive testing.

## Introduction

In non-destructive testing of elastic waveguide structures such as rods, plates, shells and beams, ultrasonic guided wave detection has the advantages of convenient excitation, long propagation distance, high sensitivity to defects and low energy consumption [1,2].Especially for non-destructive testing in significant areas such as railway transportation, oil pipelines, aircraft airframe and wings [3], the high efficiency and high precision of ultrasonic guided wave detection are more important. Therefore, using guided waves for defect detection and reconstruction has been investigated by many researchers. As early as the beginning of this century, Rose [2] clarified that ultrasonic guided waves can be used to detect pores, weak cohesion and delamination, and have considerable reliability. Eremin[4] studied the Lamb wave properties and its changes during cyclic loading of CFRP sandwich panels with aluminium honeycomb core, and based on Lamb wave, the fatigue failure and failure of the two specimens under tensile-compressive loading were identified. Puthillath[5] developed a detection method of ultrasonic guided wave linear scanning, also known as G-scan, which can detect the bonding damage of the patch during the repair of the aircraft shell, such as adhesive and cohesive weaknesses similar to that found in adhesively bonded joints. Wang et al.[6] used the Born approximation to replace the total field near the defect which is regarded as a weak scattering source with the incident field, derived the mathematical relationship between the reflection coefficient located in the far field and the defect shape function into Fourier transform pairs, and reconstructed the thinning defect in the two-dimensional plate. Sikdar[7] used probabilistic damage detection algorithm, combined with ultrasonic guided waves and surface-bonded piezoelectric wafer transducers(PWTs), to identify the location and size of the disbond and high-density core region in a honeycomb composite sandwich structure(HCSS). Da [8] et al. proposed a novel method (QDFT) for the quantitative reconstruction of pipeline defects based on ultrasonic guided SH-waves. This method started from the boundary integral equation and the Fourier transform of wavenumber domain, and derived the defect shape function using the Born approximation. Finally, the unknown defect was reconstructed through the reference model.

In recent years, many scholars have made valuable exploration and research on the application of guided waves in the field of non-destructive testing, and identified their application value. However, due to the coupling of various modes in the guided wave scattering field, it is difficult to realize high accuracy and efficiency defect reconstruction using guided wave scattering theory.[1] In addition, due to the existing defect detection and reconstruction technologies need to cooperate with the signal processing system, the actual measurement is inevitably affected by environmental noise, which will also affect the accuracy of defect reconstruction results. Therefore, considering the application of deep learning algorithms in the field of image reconstruction, an novel quantitative defects reconstruction method using the integration of data-driven method with the guided wave scattering analysis is proposed, attempts to use deep leaning algorithm to reduce the inaccuracy of the theoretical modelling and eliminate the impact of noise pollution in the process of inspection.

Since Hinton [9] made a breakthrough in the field of machine learning in 2012, deep learning has been rapidly developed and widely applied in many fields, and has demonstrated strong capabilities. In the field of image reconstruction, deep learning algorithms have many applications. In order to solve the ill-posed problem in the computational tomography process, Jin et al. [10] combined the deep convolutional neural network with the traditional inverse problem solving method in X-ray tomography, the back projection algorithm (FBP). First use the back projection algorithm to process the sub-sampled sinogram to obtain a preliminary reconstructed image, and then input the reconstructed image into the convolutional neural network for post-processing, and output a high-quality reconstructed image. In order to solve the problem of multiple scattering in image reconstruction, Yu et al. [11] divided the scattering inversion process into two steps: first, a theoretical model was used to design a back propagation algorithm, and the algorithm was used to transform the data in the measurement domain into the image domain. Then, a deep convolutional neural network with U-net structure is designed as a scattering decoder to complete the reconstruction task of image domain data. The study found that this deep learning-based image reconstruction method has higher computational efficiency and reconstruction quality than other methods when dealing with multiple scattering problems. David et al. [12] studied the combination of FBP algorithm and PWLS iterative algorithm with convolutional neural network to reconstruct images. The study found that the local fusion between this algorithm can improve the balance between resolution and variance in the image reconstruction process, so it can improve the quality of the CT image; at the same time, two different types of image reconstruction methods are used for this study (In the field of image reconstruction, FBP is a typical algorithm for directly negating forward operators, and PWLS is a typical iterative negation algorithm) illustrates the universality of local fusion of these algorithms. If the reconstruction algorithm changes, by modifying the subsequent neural network structure and then retraining, the purpose of improving the quality of the reconstructed image can still be achieved. In the review article of Michael [13], a lot of research work using deep learning algorithms for scattering inversion is listed, and it is explained that in the field of image scattering inversion, due to the lack of sample data, the mainstream method of deep learning algorithm for scattering inversion is to combine the traditional reconstruction algorithm with the deep learning algorithm. First use traditional theoretical methods for pre-reconstruction, and then input the reconstruction results into the trained machine learning model for post-processing, and finally obtain high-quality reconstruction results.

For the ultrasonic guided wave defect reconstruction, which also belongs to the inverse scattering problem, referring to the application of deep learning algorithms, especially the convolutional neural network algorithm in the field of image reconstruction, this paper proposes a defect reconstruction method combining the existing theoretical model of guided wave defect reconstruction with deep learning algorithm. By means of data-driven and theoretical analysis, high accuracy and efficiency quantitative defects reconstruction is realized.

# 1 Theoretical analysis of defect reconstruction of ultrasonic guided waves

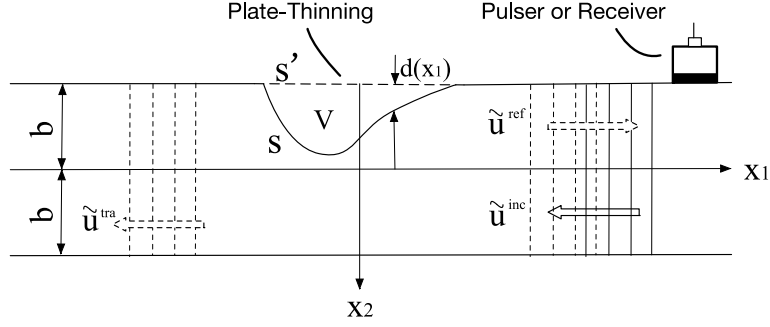

Fig.1 Reflection and transmission of an incident guided SH-wave by a plate thinning.

For the reconstruction problem of plate thinning as shown in Fig.1, the ultrasonic guided SH-wave can be excited on the right side of the plate, then the reflection coefficient can be calculated from the reflected wave signal, and then the inverse Fourier transform of the reflection coefficient in the wavenumber domain to the spatial domain can reconstruct the shape of the defect [6]. The main steps can be summarized as follows:

Starting from the wave equation in the plate and the corresponding boundary conditions, the displacement field distribution in the plate is derived. Assuming that the incident guided SH-wave in this problem is a simple mode propagating from right to left, the displacement field of the incident and reflected waves can be expressed as:

$$\tilde{u}^{inc} = A_n^{inc} f_n(\beta_n x_2) e^{+i\xi_n x_1} \qquad \tilde{u}^{ref} = A_n^{ref} f_n(\beta_n x_2) e^{-i\xi_n x_1} \qquad (1)$$

The reflection coefficient is defined as the ratio of the two coefficients:

$$C^{ref} = A_n^{ref}/A_n^{inc} \qquad (2)$$

Then through the reciprocal theorem of dynamics [14], combined with the Green function $\tilde{U}(x,X)$, in the plate, the boundary integral equation can be established:

$$\tilde{u}^{sca}(x) = \int_S \left[ \tilde{u}^{tot}(X)\tilde{T}(X,x) - \mu \frac{\partial \tilde{u}^{tot}(X)}{\partial n(X)} \tilde{U}(X,x) \right] ds(X) \qquad (3)$$

According to the defect boundary is a free boundary, $\mu \partial \tilde{u}^{tot}/\partial n = 0$. Assuming that the unknown defect is a weak scattering source, the Born assumption is introduced and the total wave field $\tilde{u}^{tot}(X)$ in equation (3) is replaced by the incident wave field $\tilde{u}^{inc}(X)$, we obtain:

$$\tilde{u}^{sca}(x) \approx \int_S \tilde{u}^{inc}(X) \mu \frac{\partial \tilde{U}(X,x)}{\partial n(X)} ds(X) \qquad (4)$$

Use Gauss theorem to convert the surface integral of the defect into the volume fraction of the void in the defect:

$$\tilde{u}^{sca}(x) \approx \int_V \left[ -k^2 \tilde{u}^{inc}(X) \mu \tilde{U}(X,x) + \mu \frac{\partial \tilde{U}(X,x)}{\partial X_i} \frac{\partial \tilde{u}^{inc}(X)}{\partial X_i} \right] dV(X) \qquad (5)$$

Substituting the Green function formula (see reference [6] for the specific derivation process), the reflected wave at the far field can be obtained:

$$\tilde{u}^{ref}(x) = \frac{i}{2b} A_n^{inc} \int_V \frac{\xi_n^2 + k^2 \cos(2\beta_n X_2)}{\xi_n} e^{2i\xi_n X_1} dV(X) \times \cos(\beta_n x_2) e^{-i\xi_n x_1}$$

(6)

Comparing equation (1) and equation (6), it can be found that the integral term in equation (6) corresponds to the reflection coefficient, and the volume fraction is expressed as a double integral, then we obtain:

$$C^{ref} = \frac{A_n^{ref}}{A_n^{inc}} = \frac{i}{2b} \frac{\xi_n^2 + k^2}{\xi_n} \int_{-\infty}^{+\infty} d(X_1) e^{2i\xi_n X_1} dX_1 \tag{7}$$

Where $C^{ref}$ is the reflection coefficient and $d(X_1)$ is the shape function of the defect. From this formula, it can be seen that $C^{ref}$ and $d(X_1)$ form a pair of Fourier transform pairs, and the defect can be reconstructed by inverse Fourier transform of $C^{ref}$.

To derive formula (7), some approximate methods are used. For example, it has been mentioned that the thinning defect is assumed to be a weak scattering source $(d \ll b)$, and then Born approximation is introduced to approximate the total field near the defect to the incident field; When calculating the Green function in the bounded plate, the incident term of the Green function at the far field is also approximately 0. These approximation methods can help simplify the theoretical model of defect reconstruction and improve the efficiency of reconstruction, but it is also inevitable to introduce model errors for the results of reconstruction and reduce the accuracy of reconstruction. Therefore, this paper proposes a new method of quantitative reconstruction of guided wave defects that combines the wavenumber spatial domain transform(WNST) and the convolutional neural network in deep learning. The neural network trained with the sample data realizes the quantitative reconstruction of defects, which not only eliminates the effects of model errors and environmental noise, but also improves the accuracy of quantitative reconstruction of defects.

## 2 Integration of theoretical analysis and data-driven methods

In order to reduce the inaccuracy of the theoretical modelling and eliminate the impact of noise pollution in the process of inspection, we consider introducing data-driven method to improve the accuracy of guided wave defect reconstruction. In this paper, we integrate the wavenumber spatial transformation(WNST) mentioned above with convolutional neural network(CNN) in the manner of local fusion. The neural network trained with the sample data can be applied to quantitatively assess and characterize defect profiles in structures.

### 2.1 Convolutional neural network

The convolutional neural network in deep learning is a deep feedforward neural network with characteristics such as weight sharing and sparse connectivity [15]. For some problem, convolutional neural networks have their unique advantages compared to traditional fully connected neural networks. For example, when processing high-dimensional data such as image signals, there are fewer training parameters for convolutional neural networks, which can improve training efficiency and make it easier to avoid overfitting [16]. In addition, objects in two-dimensional images or one-dimensional signals often have the feature of local invariance [17],which appears as scale scaling, translation, rotation and other operations that would not affect the semantic information of the object. In such problems, convolutional neural networks are easier to extract these local invariant features than fully connected feedforward neural networks.

Compared with the Backpropagation neural network, convolutional neural network contains three types of neuron connection layers: convolution layer, pooling layer and fully connected layer [18-20], as shown in Fig. 2. The main function of the convolutional layer is to perform feature extraction, and to extract the underlying features that are invariant in the image or one-dimensional signal by training the convolution kernel. As the object studied in this paper is a one-dimensional defect shape signal, the convolution kernel in the constructed convolutional neural network is also a one-dimensional numerical sequence. The main function of the pooling layer is to realize the dimensionality reduction operation of the signal while retaining the characteristic information in the signal, thereby reducing the complexity of the network and improving the calculation efficiency. The fully connected layers in convolutional neural networks are mainly distributed in the second half of the network and are often used as the output layer of the network. By controlling the structure of the output layer, the functions of the entire neural network can be changed.

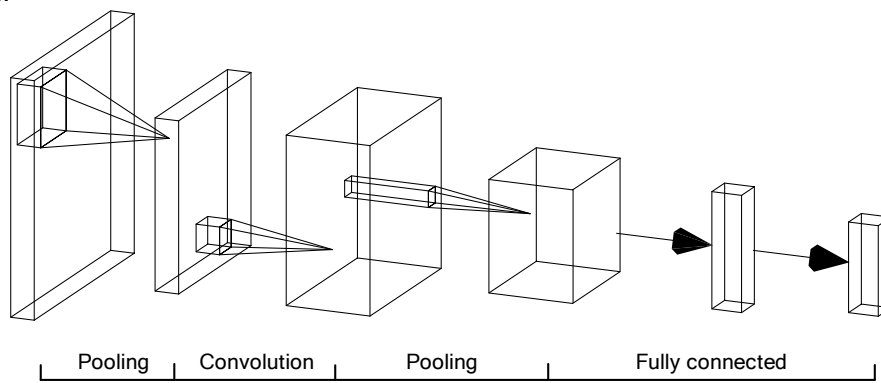

Fig.2 Typical convolutional neural network structure diagram.

The neural network constructed in this paper sets the output layer to a multi-dimensional neuron structure that can output continuous values, so as to realize the function of outputting one-dimensional defect shape signals. After completing the construction of the above neural network, give an initial value to the weight parameters in the network, and the network can realize the computing function. The training process of the network is to update the weight parameters in the network through iterative operations, so that the network can achieve a certain function. In this paper, the mean square error (MSE) is used as the performance function to evaluate the quality of the reconstruction defects calculated by the neural network. The training process is to obtain the minimum performance function value by adjusting the weight parameter value. The layered structure of the neural network determines that it has the advantage of facilitating adjustment of parameters in the network. The chain rule is used to derive inward layer by layer, and then optimize each parameter [21]. There are also various optimization algorithms in the network training process, such as stochastic gradient descent, conjugate gradient descent, Newton method, etc. [22]. In this paper, using Tensorflow[23] deep learning framework to develop a suitable neural network architecture for quantitative reconstruction of defects.

In summary, the reasons for choosing the convolutional neural network for this research are as follows: 1. The research object is a plate thinning defect with multiple geometric shapes. Using the convolutional neural network can better extract the invariant features in these geometric shapes; 2. For processing high-dimensional defect signals, the computational efficiency of convolutional neural networks is high; 3. Research on convolutional neural network algorithms is relatively mature, and a variety of network structures, performance functions, and optimization algorithms are available. It is helpful for the research of defect reconstruction methods based on deep learning.

**2.2 Data-driven-assisted defect reconstruction method(WNSTConvNet)**

The physical process of using ultrasonic to detect defects is: in the process of sound waves propagating along the medium, scattering will occur when encountering defects, resulting in transmission wave field or reflection wave field. By using the defect information contained in the transmitted or reflected waves, defect detection or reconstruction can be achieved. Therefore, guided wave defect reconstruction can be attributed to a scattering problem. For the scattering problem, it can be simply expressed by the following formula:

$$y = Tx + \xi \tag{8}$$

Where $x$ represents the scattering source, in this paper, it represents the thinning defect in the board; $y$ represents the scattering field signal; $T$ is an operator, and the properties of the operator $T$ depend on the specific scattering problem; $\xi$ is the error. The task of scattering inversion is to calculate $x$ based on $y$. The traditional main ways to solve this problem are divided into two types. The first way is to solve directly, that is, to construct the inverse problem model directly. The wavenumber spatial transformation method mentioned above belongs to this way. The expression of such methods is:

$$x = \hat{T}^{-1} y \tag{9}$$

Where $\hat{T}^{-1}$ is the theoretical reconstruction model. The advantage of this method is that for the reconstruction of defects in simple structures, the calculation of the scattering inversion can be completed in a short time; The disadvantage is that because scattering inversion is an ill-posed problem, it is difficult to calculate accurate results. In particular, when the scattering problem becomes complex, it will be extremely difficult to construct the reconstructed model, and the reconstruction accuracy will be further affected.

The second type of traditional method for solving the scattering inversion problem is an iterative-based method, such as the QDFT [9], which is expressed as:

$$O\{y\} = \arg\min_{x} f(T\{x\}, y) \tag{10}$$

The function $f$ in the formula is used to characterize the error between $T\{x\}$ and $y$. The advantage of the iterative-based method is that it can obtain accurate results; the disadvantage is that the iterative process requires a lot of calculations, and the time cost is higher.

The third method to solve the inverse scattering problem is the approach based on machine learning, that is, an inverse problem model is constructed through sample training, which can be expressed as:

$$L = \arg\min_{\theta} \sum_{n=1}^{N} f(x_n, L_\theta\{y_n\}) + g(\theta) \tag{11}$$

Where $x_n$ and $y_n$ represent a pair of training samples, the combination is expressed as ($x_n, y_n$); $N$ represents a total of $N$ pairs of training samples; $L_\theta$ is the neural network built for inversion calculations, $\theta$ is the parameter in the neural network, and it is the iterative update object during the training process; $f$ is the error function, used to characterize the difference between samples $x_n$ and $L_\theta\{y_n\}$; $g$ is a regularization term, which limits the value of parameter $\theta$ to reduce the complexity of the trained model $L_\theta$ and prevent over fitting. The advantage of using machine learning to solve the scattering inversion problem is that after training is completed, it can achieve high reconstruction accuracy and very fast calculation speed; The disadvantage is that the sample data used in the training network is difficult to obtain and the training process is also complex.

In this paper, in order to make full use of the existing defect reconstruction theory, the theoretical model (WNST) and the machine learning methods are integrated in the manner of local fusion to solve the ultrasonic guided wave defect reconstruction problem. This novel method is named WNSTConvNet. The new method is expressed as follows:

$$L = \arg\min_{\theta} \sum_{n=1}^{N} f(x_n, L_\theta\{\hat{T}^{-1} y_n\}) + g(\theta) \tag{12}$$

It can be seen from eq.(12) that in the new method proposed in this paper, the training sample pair is $(x_n, \hat{T}^{-1} y_n)$, where $x_n$ is the exact defect and $\hat{T}^{-1}$ is the theoretical construction model, $\hat{T}^{-1} y_n$ represents the pre-reconstruction defect. Therefore, we use the reconstruction result of guided wave scattering theoretical model as the training samples of the machine learning model. In the study, the mean square error(MSE) is selected as the performance function $f$ in the optimization process. The regularization term $g(\theta)$ in equation (12) selects the L2 regularization function. The neural network architecture and training process designed in this paper are shown in Fig.3.

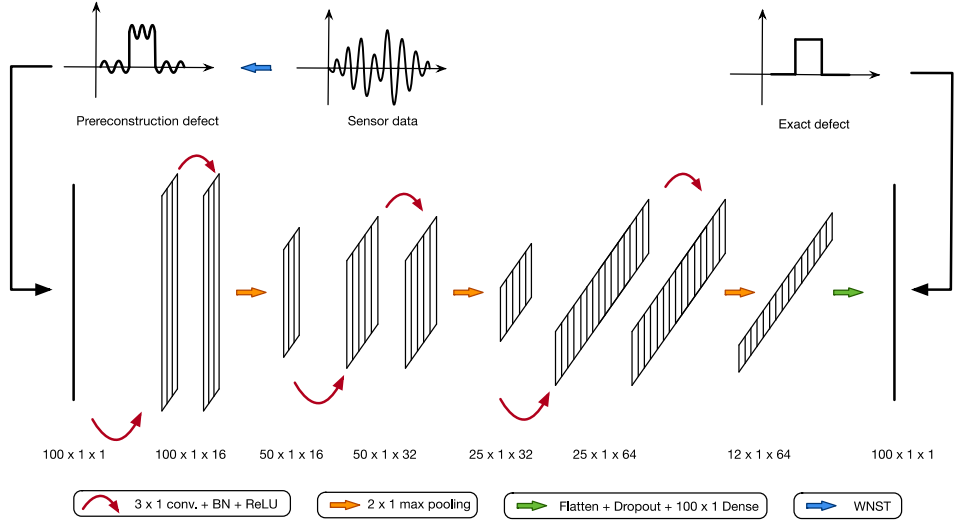

Fig.3 Convolutional neural network architecture and training process.

It can be seen from Fig.3 that the deep learning neural network processes one-dimensional signals, so a one-dimensional convolutional neural network is constructed. The training process of the neural network is: After the sensor data is processed by the wavenumber spatial domain transformation(WNST), the pre-reconstructed defects $\hat{T}^{-1}y_n$ are obtained and used as the input of the neural network. Then calculating the mean square error of the output of network with the exact defects $x_n$ to update the undetermined parameter $\theta$ in the network until the average mean square error value on the entire sample set tends to converge. When the training is completed, put a new defect signal $\hat{T}^{-1}y_n$ into the neural network, then we can obtain high-quality reconstructed defect.

In this network, ReLU [24] function is used for each convolutional layer. In order to solve the problem of gradient disappearance encountered during training, batch normalization is performed before the activation[25] to improve training efficiency; In order to solve the problem of overfitting, a Dropout layer is added at the end of the network [26] to discard some training parameters and improve the robustness of the network. At the same time, L2 regularization terms are added to limit the training parameters and improve the generalization performance of the network.

## 3 Experimental validation

In this paper, two sets of sample data are produced for training neural networks, then the trained neural networks are used for defect reconstruction testing.

### 3.1 Data preparation

First is a mixed defect datasets, which contains 1200 sets of randomly-sized isosceles triangle defects, rectangular defects, and stepped defects. Each set of sample contains two parts, one is the input sample $\hat{T}^{-1}y_n$, and the other is the reference defect $x_n$. Each part is a vector with the dimension of 100×1, and the value in the vector represents the depth of the defect. The input samples of the convolutional neural network are obtained through simulation. First, the reflection coefficient $C^{\text{ref}}$ corresponding to the exact defect is obtained by guided wave scattering theoretical analysis, and then the shape function $d(X_1)$ of the defect is calculated through the wavenumber spatial transformation, which is the required input sample. The reference defect $x_n$ is the actual defect. Among 1200 sets of sample data, 900 sets of samples are used for network training, 210 sets of samples are used for verification during the training process, and 90 sets of samples are used for the test after the training is completed.

Then is the noisy datasets. On the basis of the first training sample set , the signal strength is characterized by calculating the effective power of the input sample, and then according to the effective power, the full-band Gaussian white noise with a signal-to-noise ratio of 15dB is added to simulate the actual detection system affected by environmental noise. There are a total of 400 noisy rectangular defects in this sample set.

### 3.2 Experimental results

After the neural network training is completed, numerical examples including basic defect profiles, the stepped defect, a mixed-type defect pattern and a defect with the noisy fringe have been examined . In order to quantify the difference between the reconstructed defect and the real defect, that is, the quality of the reconstructed defect, the signal-to-noise ratio(SNR) [27] is defined:

$$SNR(\boldsymbol{x},\hat{\boldsymbol{x}}) \triangleq \max_{a \in \mathbb{R}} \left\{ 10 log_{10}(\frac{\|\boldsymbol{x}\|_{l2}^2}{\|\boldsymbol{x} - a\hat{\boldsymbol{x}}\|_{l2}^2}) \right\} \tag{13}$$

Where $x$ is the real defect, and $\hat{x}$ is the reconstructed defect. A higher SNR value corresponds to a better reconstruction. Note that the vector $x$ or $\hat{x}$ used to characterize the defect shape in this study is actually the spatial distribution of the defect shape in the entire detection range, including the defect area and the defect-free area. The purpose of this design is to not only study the quality of reconstruction defects in the defect area, but also study the noise and error in the non-defect area of the reconstruction result.

### 3.2.1 Mixed defect datasets

Using the convolutional neural network trained with the mixed defect datasets to reconstruct the triangle defects, rectangular defects and stepped defects. The testing results for the above three types of defects are shown in Fig.4, and the SNR values calculated during the testing are shown in Table 1. From the chart we can see that the WNSTConvnet method has great defect reconstruction ability. Especially for rectangular defects and stepped defects, the SNR value reached about 28dB. The average SNR value of the reconstruction results of the entire testing datasets is 23.95dB, which is nearly 200% higher than the reconstruction accuracy of the WNST method.

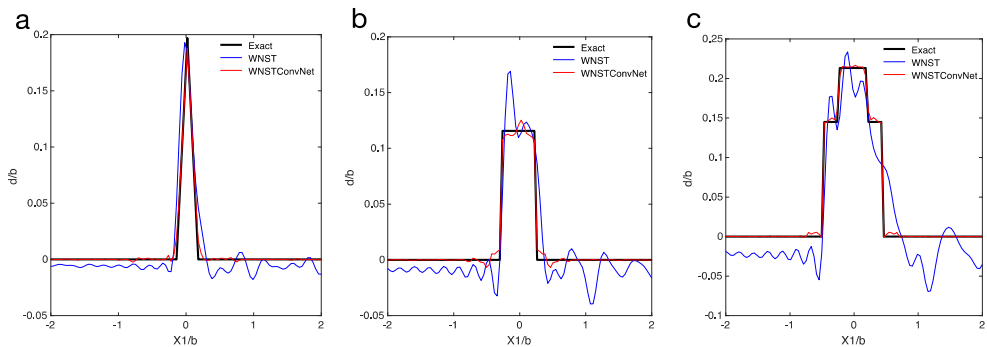

Fig.4　Reconstruct triangle defect (**a**), rectangular defect (**b**) and stepped defect (**c**) using neural networks

trained on mixed defect datasets.

Table 1 Comparison of SNR (dB) values of reconstruction results of the two methods

| Reconstruction Methods | Triangle defects | Rectangular defects | Step defects | Average accuracy |
|---|---|---|---|---|
| WNST | 9.25 | 8.13 | 7.88 | 8.20 |
| WNSTConvNet | 20.29 | 28.40 | 28.03 | 23.95 |

### 3.2.2 Noise contained datasets

In order to verify that the convolutional neural network in this study has the function of adaptive noise removal, a rectangular defect sample set containing noise is designed. By adding Gaussian white noise to the pre-reconstruction defect to simulate the random environmental noise in the actual detection process, and then input the noisy pre-reconstruction defect into the constructed convolutional neural network to remove the environmental noise. The neural network's processing results of noisy defect is shown in Fig 5, and the average SNR value on the entire test set is shown in Table 2. According to the chart, it can be seen that the convolutional neural network after training can effectively remove the 15dB noise contained in the pre-reconstruction defect. Comparing the two methods, the accuracy of the WNST is only 6.62dB, indicating that the addition of noise leads to a lower accuracy of defect reconstruction, while the result of the WNSTConvNet defect reconstruction method has an accuracy of 23.56dB, indicating that this neural network has better robustness.

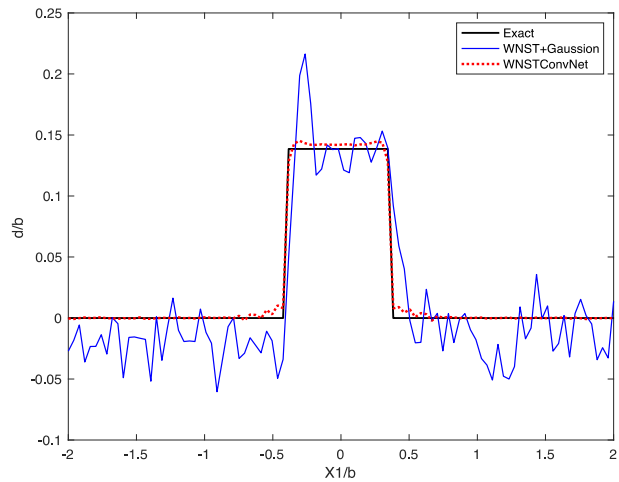

Fig.5 Reconstruction results of rectangular defects with gaussian white noise.

Table 2 SNR(dB) values of reconstruction results of rectangular defects with gaussian white noise

| Reconstruction Methods | Average accuracy |
|---|---|
| WNST | 6.62 |
| WNSTConvNet | 23.56 |

## 4 Discussion

Experimental results prove the effectiveness and robustness of the WNSTConvNet defect reconstruction method. Compared with the WNST method based on the guided wave scattering theory, WNSTConvNet which integrating data-driven model with the theoretical analysis has greater performance in the detailed processing of defect reconstruction, and the reconstructed result is closer to the real defect shape. This is of great significance to the field of high-precision defect detection in engineering. At the same time, the great robustness of the WNSTConvNet is reflected in the effective removal of samples mixed with noise in the defect reconstruction system. On the one hand, it can further improve the quality of reconstructed defects. On the other hand, removing the noise in the defect-free area is of great significance to accurately locate the defect. When using the WNSTConvNet to reconstruct defects, after the network training is completed, the time to perform a reconstruction operation is less than 1 second, which has a high reconstruction efficiency.

The drawback of the defect reconstruction method based on supervised learning is that the used network architecture can only work on information that is provided in the initial guess and the information that is extracted from the training data. For example, the neural network trained using the triangular datasets has a poor effect when reconstructing rectangular defects. According to the first experimental result, one of the solutions to this problem in practical applications is to train the neural network with datasets of a variety of typical geometrical information. In addition, you can add a classifier before the reconstruction model, and then input different types of pre-reconstruction defects into the corresponding convolutional neural network for computational reconstruction. Another shortcoming of using neural networks to reconstruct defects is the need to prepare a large amount of training data. At present, the amount of relevant training data in engineering is small and the cost of obtaining data through experiments is high. Using computer simulation to obtain data to train a neural network and then use the trained network in practice is a feasible method to solve data problems.

Data-driven-assisted defect reconstruction method proposed in this paper does not specify the type of integrated theoretical model and the machine learning model. Wavenumber spatial transformation(WNST) and convolutional neural network are selected as subjects of study in this paper to prove the effect of this method.

## 5 Conclusion

This paper proposes a novel data-driven-assisted ultrasonic guided wave defect quantitative reconstruction method (WNSTConvNet), which integrates the wavenumber spatial transformation method(WNST) with a convolutional neural network in a local fusion manner. The neural network architecture is a one-dimensional convolutional neural network. The L2 regularization term and the Dropout layer are added to prevent overfitting during training, and batch normalization is performed in each convolutional layer to prevent the vanishing gradient.

Through experiments, the reconstruction results of WNSTConvNet are compared with those of the wavenumber spatial transformation method, which verifies that the WNSTConvNet method is effective and has higher reconstruction accuracy and stability. The neural network trained with mixed datasets has a reconstruction accuracy of 20dB for the three types of defects, which shows that the algorithm has good generalization performance. Especially, for the reconstruction of rectangular defects and stepped defects, the results by the proposed method is nearly 200% more accurate than the solution by the WNST method. In addition, considering the combined defect with noisy fringe, WNSTConvNet method can achieve the purpose of removing noise, which shows that this method has good robustness.

This method has a high calculation speed when reconstructing defects. For the SH-waves reconstruction problem of plate thinning defects studied in this paper, the defect reconstruction calculation can be completed within 1 second. So it's a high-precision and high-efficiency defect quantitative reconstruction method compared with the analytical method and provides a valuable insight into the development of artificial intelligence-assisted inspection systems with high accuracy and efficiency in the field of non-destructive testing.

The drawbacks of this method: The quality of reconstruction results depend on the information from training data and the initial results. If the researched problem exceeds the applicable scope of the trained network, it is necessary to reacquire sample data and train a new network in a targeted manner. In addition, deep learning algorithms have high requirements for the quality and quantity of training data, but in actual engineering applications, it is difficult to obtain a large amount of effective data. In response to these problems, the follow-up work will do further research on improving the universality of the WNSTConvNet method in application and solving data problems.

**References**
[1]Su, Z.; Ye, L.; Lu, Y. Guided Lamb waves for identification of damage in composite structures: A review[J]. Journal of Sound and Vibration **2006**, 295, 753-780.

[2]Rose, J.L. A baseline and vision of ultrasonic guided wave inspection potential[J]. Journal of Pressure Vessel Technology **2002**, 124, 273.

[3] Qiu, L.; Yuan, S.; Mei, H.; Fang, F. An improved Gaussian mixture model for damage propagation monitoring of an aircraft wing spar under changing structural boundary conditions[J]. *Sensors* **2016**, *16*, 291.

[4]Eremin, A.V.; Burkov, M.V.; Byakov, A.V.; Lyubutin, P.S.; Panin, S.V.; Khizhnyak, S.A. Investigation of acoustic parameters for structural health monitoring of sandwich panel under cyclic load. *Key Engineering Materials* **2016**, *712*, 319-323.

[5]Puthillath, P.; Rose, J.L. Ultrasonic guided wave inspection of a titanium repair patch bonded to an aluminum aircraft skin. *International Journal of Adhesion and Adhesives* **2010**, *30*, 566-573.

[6]Wang B, Hirose S. Inverse problem for shape reconstruction of plate-thinning by guided SH-waves[J]. The Japanese Society for Non-Destructive Inspection, 2012, 53(10): 1782-1789.

[7]Sikdar, S.; Banerjee, S. Identification of disbond and high density core region in a honeycomb composite sandwich structure using ultrasonic guided waves. *Composite Structures* **2016**, *152*, 568-578.

[8] Da Y , Dong G , Wang B , et al. A novel approach to surface defect detection[J]. International Journal of Engineering Science, **2018**, 133(DEC.):181-195.

[9] Hinton, G.E.; Srivastava, N.; Krizhevsky, A.; Sutskever, I.; Salakhutdinov, R.R. Improving neural networks by preventing co-adaptation of feature detectors[J]. Computer Science **2012**, arXiv: 1207.0580.

[10] Jin K H , Mccann M T , Froustey E , et al. Deep Convolutional Neural Network for Inverse Problems in Imaging[J]. IEEE Transactions on Image Processing, **2017**:1-1.

[11] Yu Sun, Zhihao Xia, Ulugbek S. Kamilov. Efficient and accurate inversion of multiple scattering with deep learning[J]. Computer Vision and Pattern Recognition. 10.1364/OE.26.014678

[12] D. Boublil, M. Elad, J. Shtok, and M. Zibulevsky, Spatially-adaptive reconstruction in computed tomography using neural networks[J]. IEEE Trans. Med. Imag., vol. 34, no. 7, pp. 1474–1485, July **2015**.

[13] M. T. McCann, K. H. Jin and M. Unser, Convolutional Neural Networks for Inverse Problems in Imaging: A Review[J]. IEEE Signal Processing Magazine, vol. 34, no. 6, pp. 85-95, Nov. **2017**, doi: 10.1109/MSP.2017.2739299.

[14] Achenbach JD. Reciprocity in Elastodynamics. Cambridge University Press, 2003.

[15] Y. Jia et al., in CAFFE: Convolutional Architecture for Fast Feature Embedding[J], **2014**, arXiv preprint arXiv:1408.5093.

[16] Y. LeCun, Y. Bengio, and G. Hinton, Deep learning[J]. Nature, vol. 521, no. 7553, pp. 436–444, May **2015**

[17] A. Krizhevsky, I. Sutskever, and G. E. Hinton, "Imagenet classification with deep convolutional neural networks[J]," in Proc. 25th Int. Conf. Neural Information Processing Systems, Lake Tahoe, NV, **2012**, pp. pp. 1097–1105.

[18] Gao Li-Gang, Chen Pai-Yu, Yu Shi-Meng. Demonstration of convolution kernel operation on resistive cross-point array[J]. IEEE Electron Device Leters, **2016**, 37(7):870-873

[19] Gu Jiu-Xiang，Wang Zhen-Hua, Jason Kuen, et a1. Recent advances in convolutional neural networks[J]. arXiv:1512. 07108v5, **2017**

[20] Sainath T N , Mohamed A , Kingsbury B, et a1. Deep convolutional neural networks for LVCSR[J]//Procedings of the IEEE International Conference on Acoustics, Speech and Signal Processing. Vancouver，Canada，**2013**:8614—8618

[21] LeCun Y, Botou L, Bengio Y, et a1. Gradient-based learning applied to document recognition[J]. Proceedings of the IEEE, **1998**, 86(11):2278-2324

[22] Duchi J, Hazan E, Singer Y. Adaptive subgradient methods for online learning and stochastic optimization [J] . Journal of Machine Learning Research，**2011**，12 (7):257-269.

[23] Abadi, M. *et al.* TensorFlow: large-scale machine learning on heterogeneous distributed systems[J]. Preprint at https://arxiv.org/abs/1603.04467 (**2016**).

[24] Nair V, Hinton G E, Farabet C. Rectified linear units improve restricted Boltzmann machines[J]//Proceedings of the 27th International Conference on Machine Learning. Haifa, Israel, **2010**:807-814

[25] Ioffe S，Szegedy C. Batch normalization: Accelerating deep network training by reducing internal covariate shift[J]. arXiv:1502.03167, **2015**

[26] Yoo H-J. Deep convolution neural networks in computer vision: A review[J] . IEIE Transactions on Smart Processing and Computing，**2015**，4(1):35-43

[27] D. Boublil, M. Elad, J. Shtok and M. Zibulevsky, "Spatially-Adaptive Reconstruction in Computed Tomography Using Neural Networks[J]," in IEEE Transactions on Medical Imaging, vol. 34, no. 7, pp. 1474-1485, July **2015**, doi: 10.1109/TMI.2015.2401131.